\documentclass[superscriptaddress,twocolumn]{revtex4}
\usepackage{bm}
\usepackage{graphicx}
\usepackage{hyperref}

\begin{document}

\title{Electronic states in nanowires with hexagonal cross-section}

\author{I. A. Kokurin}
\email[E-mail:]{kokurinia@math.mrsu.ru} \affiliation{Institute of
Physics and Chemistry, Mordovia State University, 430005 Saransk,
Russia} \affiliation{Ioffe Institute, 194021 St. Petersburg, Russia}
\affiliation{St. Petersburg Electrotechnical University ``LETI'',
197376 St. Petersburg, Russia}

\begin{abstract}
The electron spectrum in a uniform nanowire with a hexagonal
cross-section is calculated by means of a numerical diagonalization
of the effective-mass Hamiltonian. Two basis sets are utilized. The
wave-functions of low-lying states are calculated and visualized.
The approach has an advantage over mesh methods based on
finite-differences (or finite-elements) schemes: non-physical
solutions do not arise. Our scheme can be easily generalized to the
case of multi-band (Luttinger or Kane) ${\bf k}\cdot{\bf p}$
Hamiltonians. The external fields (electrical, magnetic or strain)
can be consistently introduced into the problem as well.
\end{abstract}

\date{\today}

\maketitle

\section{Introduction}

Semiconductor nanowires (NWs) are currently of great interest due to
the possibility of their application in electronics. NWs can be used
as a work item of field-effect
transistors~\cite{Dayeh2007,Chuang2013},
photodetectors~\cite{Dai2014}. Moreover, there is a wide variety
of NW-based photonic devices including light-emitting diodes,
chemical and gas sensors, waveguides, solar cells and nonlinear
optical converters \cite{Yan2009,Jia2019}. The NW-based structures
are also of fundamental interest. The topological states of the
matter and Majorana fermions are realized in NWs due to the
proximity effect~\cite{Stanescu2011,Alicea2012,Mourik2012,Gul2018}.

Usually, NWs of III-V materials with a zinc-blende lattice are grown
in [111] crystal direction, that leads to the hexagonal shape of
NW's cross-section (Fig.~\ref{fig01}a). Early, simple models of NW
with circular or square cross-sections were used for the calculation
of the charge carrier spectrum and wave functions. However, for
optical and transport applications it is necessary to know the
carrier subband spectrum with higher precision, i.e., take into
account a real NW's shape.

The NW's translation invariance in longitudinal direction simplifies
the problem: one needs simply to solve the spectral problem for a
two-dimensional electron bounded in a hexagon. Usually, the
finite-difference (or finite elements) method is used for this
purpose \cite{Degtyarev2017,Sitek2018,Wojcik2018}. This problem is
nontrivial even for the case of the electron in a non-degenerate
band described by the scalar effective mass. We propose an
alternative approach based on the numerical diagonalization of the
matrix Hamiltonian written in an appropriate basis.

\section {Hamiltonian and basis functions}

\begin{figure}
\includegraphics[width=0.85\linewidth]{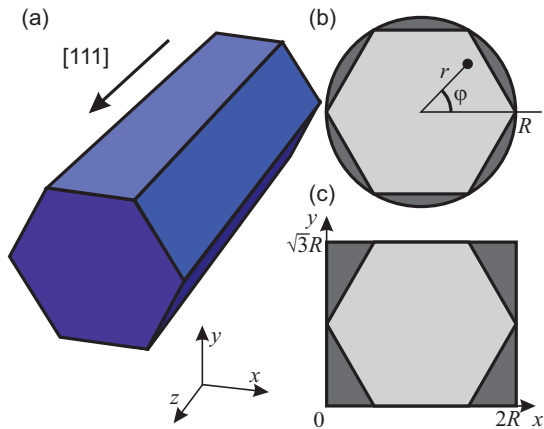}
\caption{\label{fig01} (a) The sketch of NW with a hexagonal
cross-section. (b) The use of circular basis set (\ref{basis_c}).
The shaded areas between the circle and the hexagon serve as
additional potential barriers. (c) The same as in (b) for the
rectangular basis (\ref{basis_r}).}
\end{figure}

The effective potential barrier bounding an electron in NW is equal
to the electron affinity $\chi$ (several eV). Within the effective
mass approximation, such a height is equivalent to an infinite
barrier. To find electronic states in NW with a hexagonal
cross-section we propose to use the matrix mechanics. It is
convenient to choose the eigenfunctions of the
Hamiltonian $H_0$, which describes electrons in NW with a circular
or rectangular cross-section, as the basis functions. The corresponding circle or rectangle
is chosen to be circumscribed around the hexagon (see
Fig.~\ref{fig01}b,c). The spectral problem is reduced to the problem
with the Hamiltonian $H=H_0+V({\bf r})$, where $V({\bf r})$ is
nonzero in shaded areas of Fig.~\ref{fig01}b,c. The height of this
potential cannot be chosen infinite at calculation, however, we do
not make a big mistake putting it to be finite but high, e.g.,
$V_0\sim\chi$. The envelope function approximation in a single band
with a scalar effective mass $m^*$ is used. The spin-dependent terms
are excluded from consideration.

The eigenfunctions and eigenenergies for the electron in the
infinite circular potential well of radius $R$ are well-known

\begin{equation}
\label{basis_c} \Psi^0_{mn}(r,\varphi)=\frac{\sqrt
2}{RJ_{|m|+1}(j_{mn})}J_{|m|}\left(j_{mn}\frac{r}{R}\right)\frac{1}{\sqrt{2\pi}}e^{im\varphi},
\end{equation}
\begin{equation}
E^0_{mn}=\frac{\hbar^2j_{mn}^2}{2m^*R^2},
\end{equation}
where $m=0,\pm 1,\pm 2,...$, $n=1,2,...$; $J_{m}(x)$ is the first
kind Bessel function, and $j_{mn}$ is the $n$th zero of $J_{m}(x)$.

In the case of a potential well of rectangular shape circumscribing
the same hexagon, the eigenfunctions and eigenenergies are given by

\begin{equation}
\label{basis_r}
\Psi^0_{mn}(x,y)=\frac{\sqrt{2}}{3^{1/4}R}\sin\left(\frac{\pi
mx}{2R}\right)\sin\left(\frac{\pi ny}{\sqrt 3R}\right),
\end{equation}
\begin{equation}
E^0_{mn}=\frac{\pi^2\hbar^2}{8m^*R^2}\left(m^2+\frac43n^2\right)
\end{equation}
with $m,n=1,2,...$ .

We will search for the electron wavefunctions in hexagonal NW (h-NW)
as a series in above basis sets

\begin{equation}
\label{psi} \Psi_j({\bf r})=\sum_{mn}C^j_{mn}\Psi^0_{mn}({\bf r}).
\end{equation}

The spectral problem is reduced to finding the eigenvalues of the
Hamiltonian $H=H_0+V({\bf r})$ matrix. For the matrix elements we
have, $\langle
m'n'|H|mn\rangle=E^0_{mn}\delta_{m'm}\delta_{n'n}+\langle
m'n'|V({\bf r})|mn\rangle$. The latter term is proportional to the
overlap integral $I_{m'n';mn}$ of the basis functions in the single
barrier segment.

We can use some symmetry arguments for the matrix elements
calculation. They are given by
\begin{equation}
\langle m'n'|V({\bf r})|mn\rangle=6V_0\delta_{m',m+6M}I_{m'n';mn},
\end{equation}
and
\begin{equation}
\langle m'n'|V({\bf
r})|mn\rangle=4V_0\delta_{m',m+2M}\delta_{n',n+2N}I_{m'n';mn},
\end{equation}
for the case of a circular and rectangular basis, respectively. Here
$M,N=0,\pm 1,\pm 2,...$. In the latter case the overlap integral can be
found analytically, but in the former, only numerically.

\section {Numerical diagonalization}

\begin{figure*}
\includegraphics{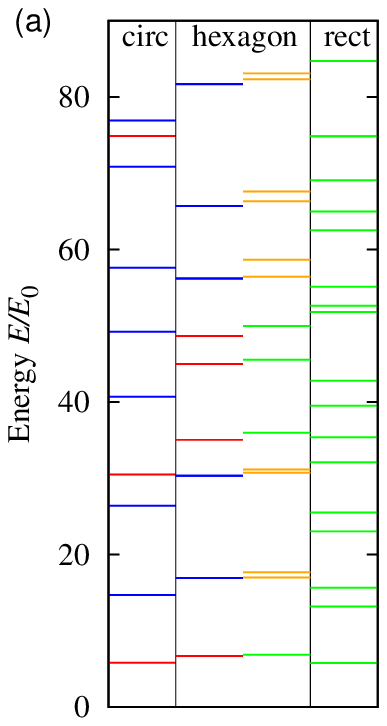}
\includegraphics{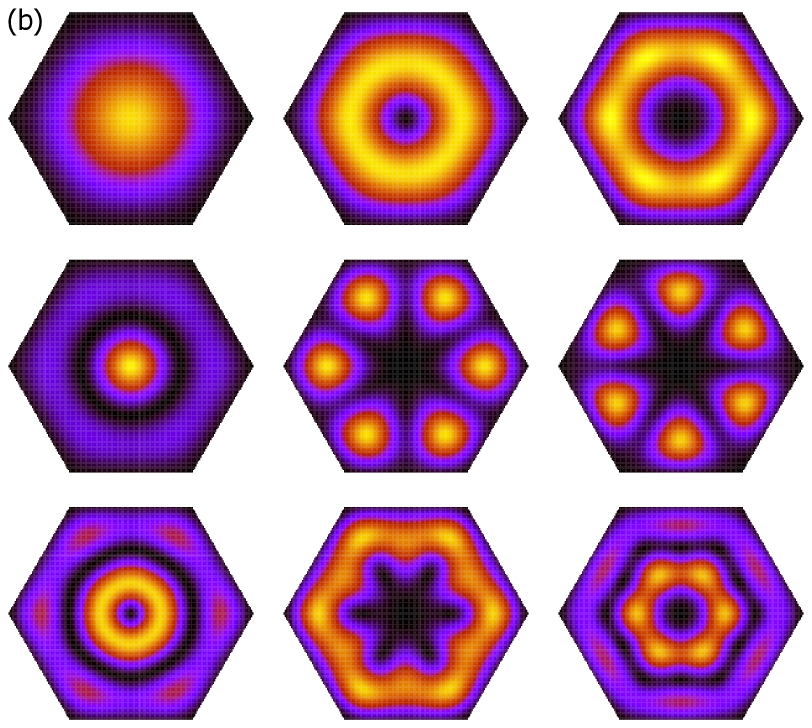}
\caption{\label{fig02} (a) Subband energies of the circular (left),
rectangular (right) and hexagonal (center) NWs. The energy levels
for h-NW are calculated using the numerical diagonalization of the
Hamiltonians in both circular and rectangular basis sets. Blue and
orange lines denote the doubly degenerate (quasi-degenerate) levels.
(b) The spatial distribution of the electron density $|\psi|^2$ in
h-NW for 1, 2(3), 4(5), 6, 7, 8, 9(10), 11(12), 13(14)-th subband.}
\end{figure*}

For the numerical diagonalization of derived matrix Hamiltonians one
needs to truncate the matrix dimension. At the same time, we have to
choose a matrix size so as to ensure acceptable accuracy. The
maximal values of $m_{max}$ and $n_{max}$ determine the size of the
truncated matrix. In the circular basis the matrix dimension is
$(2m_{max}+1)n_{max}$, while in rectangular one we have
$m_{max}n_{max}$. The position of calculated subband bottoms in h-NW
is depicted in the central section of Fig.~\ref{fig02}a. The results are
depicted for the truncated matrix of dimension $775\times 775$ and
$1000\times 1000$ for circular and rectangular bases, respectively.
This corresponds to the choice of $m_{max}=15$, $n_{max}=25$ and
$m_{max}=40$, $n_{max}=25$, respectively. The energies are scaled to
the value $E_0=\hbar^2/2m^*R^2$, that for the case of GaAs NW
($m^*=0.067m_0$ \cite{Vurgaftman2001}) with $R=20$ nm is equal to
1.41 meV. The barrier height $V_0$ was set to $10^3E_0$.

The energy levels are a single or twofold degenerate (excluding
spin). This is especially easy to trace when considering a circular
basis. In this case the degenerate states arise even at diagonalization
of the Hamiltonian matrix of small size, which does not provide a good
precision. In this sense the use of Cartesian basis is more
appropriate (there are no degenerate states) in order to track the
convergence of the method with a growing matrix dimension.
Nonetheless, the use of a Cartesian basis requires a larger matrix
size to attain the same precision as for a circular basis. Moreover,
for the case of a non-degenerate Cartesian basis, the real twofold
degeneracy of states is reached only in the limit of
$V_0\rightarrow\infty$, $m_{max},n_{max}\rightarrow\infty$. This is
due to the lack of 6-th order symmetry axis in the model described
in Fig.~\ref{fig01}c compared to that in Fig.~\ref{fig01}b.

The calculated coefficients $C^j_{mn}$ give us the opportunity to
find the spatial behavior of wave functions [see Eq.~(\ref{psi})].
The electron distributions $|\Psi|^2$ corresponding to energy levels
of Fig.~\ref{fig02}a (central panel) are depicted in
Fig.~\ref{fig02}b. The wave functions of degenerate states
calculated in the Cartesian basis, in general, do not possess
hexagonal symmetry. However, the total electron density at
degenerate levels has this property.

\section{Conclusions}

In conclusion, the electronic states in h-NW are calculated.
Corresponding wave functions are numerically found and visualized.
The given approach can be generalized onto the case of the hole
quantization in NWs or the more general case of a multi-band
Hamiltonian. In the same manner the external fields can be
introduced into the problem. The core-shell (or core-multi-shell)
structures with a hexagonal cross-section of the core and shells can
be considered by analogy with circular ones
\cite{RaviKishore2010,Rudakov2019}. This approach is attractive
because there is no need to impose boundary conditions at the
heterointerfaces, which is usually the case when calculating
electronic and hole states in heterostructures using wave mechanics
\cite{Bastard1988}. Moreover, non-physical solutions do not arise in
our approach compared to other ones.


\end{document}